\documentstyle[floats,epsf,twocolumn,eqsecnum,prb,aps]{revtex}
\newcommand{\tphi}{\widetilde\phi}
\newcommand{\no}{\stackrel{\textstyle{.}}{.}}
\newcommand{\Ymark}
{{\normalsize\boldmath $Y\hspace{-1mm}ukawa~
I\hspace{-.3mm}n\hspace{-.3mm}stitute~K\hspace{-.5mm}yoto$}
\hfill {\rm\normalsize YITP-95-17}\\~ \\}
\begin{document}
\draft
\title{\Ymark Adiabatic Ground-State Properties of Spin Chains 
with Twisted Boundary Conditions}
\author{ Takahiro Fukui \cite{JSPS}} 
\address{Yukawa Institute for Theoretical Physics, Kyoto University,
Kyoto 606-01, Japan}
\author{Norio Kawakami}
\address{Department of Applied Physics, and
Department of Material and Life Science,\\
Osaka University, Suita, Osaka 565, Japan}
\date{February 13, 1996}
\maketitle
\begin{abstract}
We study the Heisenberg spin chain with twisted boundary 
conditions, focusing on the adiabatic flow of the energy spectrum
as a function of the twist angle. In terms of effective field 
theory for the nearest-neighbor model, we show that the 
period 2 (in unit $2\pi$) obtained by Sutherland and Shastry arises 
from irrelevant perturbations around the massless fixed 
point, and that this period may be rather general for one-dimensional 
interacting lattice models at half filling. In contrast, the period  for 
the Haldane-Shastry spin model with $1/r^2$ interaction  has 
a different and unique origin for the period, namely, 
it reflects fractional statistics in Haldane's sense.
\end{abstract}

\section{Introduction}

In recent rapid progress in one-dimensional (1D) quantum systems, 
exactly solvable models have been playing a vital role to understand 
low-energy critical properties. 
For example, the idea of twisted boundary conditions 
has been successfully combined with the exact solution to 
reveal a rich structure of the 
excitation spectrum. In this context, Sutherland and Shastry 
examined the spectral flow in the $XXZ$ Heisenberg model
with twisted boundary conditions using the Bethe ansatz 
method.\cite{SutSha} They
found a remarkable feature in the spectral flow,
which is explained briefly here.
Note that the Heisenberg $XXZ$ model 
with twisted boundary conditions 
is equivalent to the interacting fermion model
on a ring threaded by a magnetic flux. 
Since each fermion on a ring feels a gauge potential, it acquires 
a non-trivial phase factor when it moves along the ring 
and returns to the previous position. Let us denote 
this phase as $e^{2 \pi i\phi}$.
Recall first that the full spectrum at $\phi =1$ should be 
equivalent to that at $\phi =0$. 
However, if we increase $\phi$ gradually from $0$ and follow each
eigenstate adiabatically, each state does not necessarily come back 
to the original state at $\phi =1$ since the spectral 
flow occurs and levels cross each other.
Sutherland and Shastry found that the ground state at $\phi =0$
becomes the first excited state at $\phi =1$, and returns to the
ground state again at $\phi =2$
for the antiferromagnetic $XXZ$ model
with zero magnetic field.  
The resulting period $2$
shows a sharp contrast to that for non-interacting systems 
which have the period $N$ for $N$-site systems.
In this  connection, ferromagnetic spin models 
were discussed in \cite{YuFow,Sut,RES,FIO},
paying attention to the motion of string solutions.
Also, Berry's phase was calculated in \cite{KorWu}. 
The spectral flow in the Hubbard model and the 
{\it t-J} model was investigated in \cite{Aoki}.

Stimulated by the above studies, we are naturally lead to
fundamental questions, i.e. what is essential to determine
the period in the spectral flow, and how the 
spectral flow occurs for more generic cases. 
In this paper we reexamine this problem. We show
that the presence of irrelevant perturbations
among the massless fixed point is essential to determine the
period of $2$, implying that the period $2$ may be 
rather general for 1D interacting systems with massless
excitations
at half-filling (or spin systems
with zero magnetic field). 
We find, however, that
this is not the case for a special class of models, 
i.e. the 1D models with $1/r^2$ interaction. 
One can see a different origin for the periodicity
in this model, namely, the period is 
controlled by Haldane's fractional exclusion  statistics.
\cite{Hal}

This paper is organized as follows.
In the next section, we first reexamine how the 
spectral flow occurs for the antiferromagnetic Heisenberg 
$XXZ$ chain by using low-energy effective theory. 
In \S 3, we study the Haldane-Shastry model
with $1/r^2$ long-range interaction, and 
find that the period of the ground 
state is determined by the statistical interaction.
We discuss the results in terms of the motif picture
in  fractional exclusion statistics.\cite{Hal}
 A brief summary is given in \S 4.

\section{Adiabatic Flow in the {\it XXZ} Heisenberg Chain}

We first study the Heisenberg $XXZ$ spin chain model
with nearest neighbor interactions.
The corresponding Hamiltonian is  
given for the system with even $N$ sites,
\begin{eqnarray}
H&=&\sum_{j=1}^{N}
\left[ -\frac{1}{2}J(S_j^+S_{j+1}^-+h.c.)
+J_zS_j^zS_{j+1}^z\right]\nonumber\\
&=&H_0+H_{{\rm int}},
\label{Ham1}
\end{eqnarray}
where $S_j^a$ is a spin-1/2 generator at $j$th site, and $J$
and  $J_z$ are antiferromagnetic coupling constants with $J_z\ge0$.
We are concerned with the massless phase of
this model $(J_z/J \le 1)$ with twisted boundary conditions,
\begin{equation}
S_{j+N}^\pm={\rm e}^{\pm 2\pi{\rm i}\phi}S_j^\pm,
\quad S_{j+N}^z=S_j^z \quad {\rm for ~any~} j. 
\label{TwiBou}
\end{equation}
After the work of Sutherland and Shastry,\cite{SutSha}
it was found in the numerical studies of the 
Bethe-ansatz solution \cite{YuFow,RES}
that the gap formation due to the finite-size effect
may be relevant for the period of the spectral flow.
To see what kind of the interaction  is 
actually relevant, and moreover to 
address the problem in more general 
cases including non-integrable models,
we here analyze the spectral flow by means of 
effective field theory in the continuum limit.
According to the Jordan-Wigner transformation,
$S_j^z=1/2-n_j$, and $S_j^+=\psi_j{\rm e}^{{\rm i}\pi\sum_{k(<j)}n_k}$,
where $n_j\equiv\psi_j^\dagger\psi_j$, 
the Hamiltonian (\ref{Ham1}) with the boundary
condition (\ref{TwiBou}) is rewritten 
in terms of the fermion field $\psi_j$,
\begin{eqnarray}
&&H_0=-\frac{J}{2}\sum_{j=1}^{N}\left(\psi_j^\dagger\psi_{j+1}+
{\rm h.c.}\right),\label{HamFre}\\
&&H_{{\rm int}}=J_z\sum_{j=1}^{N}
\left( n_j-\frac{1}{2}\right)
\left( n_{j+1}-\frac{1}{2}\right) .
\label{HamInt}
\end{eqnarray}

Henceforth, we restrict ourselves to the half-filling case
(zero magnetic field) with even $N$.  In order to correctly 
describe the spectral flow for the finite system, 
one should seriously cope with 
the parity effect on the ground state, i.e. the effect depending on
whether the number of particles is even or odd. 
This is briefly summarized in Appendix A.
Passing to the continuum limit, the fermion operators on the lattice 
$\psi_j/\sqrt{a/2\pi}$
can be expressed in terms of  the continuum field operators,\cite{Aff}
$\psi (x)={\rm e}^{-{\rm i}p_Fx}\psi_L(x)+{\rm e}^{{\rm i}p_Fx}\psi_R(x)$
with $\psi_{L(R)}(x)=\sqrt{\frac{2\pi}{l}}
\sum_{k\in{\bf Z}}{\rm e}^{2\pi{\rm i}(k+\tphi )x/l}
\psi_{L(R),k+\tphi}$, 
where $l$ is the linear size of the system,
related with the lattice spacing $a$ by $l=aN$,
$p_F$ is the Fermi momentum defined by $p_F=\pi/2a$
and $\tphi=\phi+1/2$ (see Appendix A).
The one-particle dispersion is given by a linear form
$\epsilon (p)\sim\pm v_Fp =\pm\frac{2\pi}{l}v_F(k+\tphi )$
around the Fermi point $p=\pm p_F$, 
where $v_F$ is the Fermi velocity defined by $v_F\equiv Ja$.
Then, there appears infinitely deep Dirac sea which 
have to be  regularized. In the presence of 
twisted boundaries, this regularization is somehow subtle.
First, we define the vacuum state as
the lowest energy state at a fixed fermion
number for a given $\phi$, which is written as
\begin{equation}
|0\rangle =\prod_{k+\tphi > 0}\prod_{k'+\tphi\le 0}
\psi_{L,k+\tphi}^\dagger\psi_{R,k'+\tphi}^\dagger |0) ,
\label{Vac}
\end{equation} 
where $|0)$ denotes the reference state with no particles.
Accordingly, the Hamiltonian (\ref{HamFre}) is rewritten as 
\begin{equation}
H_0=\sum_{k\in{\rm Z}}v_Fp
\left(\no\psi_{R,k+\tphi}^\dagger\psi_{R,k+\tphi}\no -
\no\psi_{L,k+\tphi}^\dagger\psi_{L,k+\tphi}\no\right),
\label{LowHam}
\end{equation}
where normal ordering  $\no~\no$ 
for the vacuum (\ref{Vac}) is defined according to the 
sign of $k+\tphi$, not that of $k$.
The Casimir energy is easily obtained as 
$E_v=\frac{2\pi v_F}{l}\left( \Phi^2-\frac{1}{12}\right)$
by the $zeta$-function regularization, where 
$\Phi\equiv\phi$ for $0\le\phi\le 1/2$ and $ \Phi \equiv 1-\phi$ for
$1/2\le\phi\le 1$, 

Since low-energy properties of the present system can be described 
by the shifted U(1) Kac-Moody algebra for 
left- and right-going currents,
\begin{equation}
J_{L(R)}(x)=\no\psi_{L(R)}^\dagger\psi_{L(R)}\no (x)\mp
\frac{2\pi}{l}\phi ,
\end{equation}
the spectrum of the Hamiltonian (\ref{LowHam}) is given by its
representation. As is well known, there exist  infinite
number of the primary states labeled by 
integers $Q_L$ and $Q_R$, \cite{Com1} 
each of which forms a
conformal tower with descendant 
states labeled by integers $N_L$ and $N_R$.
By taking into account the effect of twisting, 
we have the spectrum for the 
eigenstate denoted by $|Q_L,Q_R;N_L,N_R; \phi \rangle$, 
\begin{eqnarray}
E_0&&(Q_L,Q_R;N_L,N_R;\phi)\nonumber\\
&&=\frac{1}{2} (Q_L-\phi)^2+\frac{1}{2}(Q_R +\phi)^2+
N_L+N_R-\frac{1}{12}
\label{Energy}
\end{eqnarray}
in unit $2\pi v_F/l$.
Since we restrict ourselves to the half-filling, we 
have the constraint for quantum numbers,  $Q_L+Q_R=0$. 

We start by examining how the spectral flow of 
(\ref{Energy}) occurs
as a function of $\phi$, paying special attention to the motion
of the ground state for $\phi=0$  which
is denoted by $ |a \rangle 
\equiv |0,0;0,0;\phi\rangle$.
This state remains as the ground state up to $\phi=1/2$ 
as seen from eq.(\ref{Energy}) and also from Fig.\ref{OnePF}.
At this point there occurs a level-crossing  
between the state $|a \rangle $ and the primary state
$|1,-1;0,0;\phi\rangle$ which is 
the first excited state at $\phi=0$.
In other words, we can say that 
there occurs a pair-creation
from the absolute ground state (lowest energy state)
at this level-crossing point,
 as drawn in Fig.1. Note that this pair creation is related to the fact 
that one of the rapidities on the real
axis jumps to the ${\rm i}\pi$ line with the twist angle being increased
in the Bethe ansatz description.\cite{YuFow,Sut,RES,FIO}
As $\phi$ is further increased, the initial ground state 
 $|a \rangle$ becomes  the 
first excited state at $\phi=1$, where there occurs
four-fold degeneracy among the state $|a \rangle$, the 
other primary state 
$|b \rangle \equiv |2,-2;0,0;\phi\rangle$,
and two descendant states 
$|c \rangle \equiv |1,-1;1,0;\phi\rangle$,
$|d \rangle \equiv |1,-1;0,1;\phi\rangle$.
At this point, the whole 
spectrum recovers the original structure at $\phi =0$,
though  the initial ground state $|a \rangle$ 
is raised up to the  excited state.
If we further increase $\phi$, the state $|a \rangle$ is  
continuously raised to highly excited 
state with  smoothly crossing with
other levels. 
This smooth rise of the ground state implies that the period of the
spectral flow becomes macroscopic, which is consistent with 
the known fact that the period of the spectral flow 
for non-interacting systems is given by  $N$ for $N$-site systems.
\cite{SutSha}

Having noticed that the period for free fermions is $N$,
we now investigate how the spectral 
flow of these states is modified when we turn on the interactions.
We wish to clarify what kind of the interaction
brings about the period of 2 
obtained by Sutherland and Shastry for 
the $XXZ$ model (interacting systems).
To this end, we bosonize the interaction term in
a standard way. Expressing the interaction 
Hamiltonian in terms of the field operator,
we write down the Hamiltonian as  $H=H_0'+H_J+H_b$, 
\begin{eqnarray}
&&H_J=4\frac{g}{2\pi}\int_0^l {\rm d}xJ_LJ_R ,\\ 
&&H_b=-\frac{g}{2\pi}\int_0^l {\rm d}x\left[
(\psi_L^\dagger\psi_R)^2+(\psi_R^\dagger\psi_L)^2\right], 
\end{eqnarray}
where $g\equiv J_za/2\pi$,
$H_0'$ is a free Hamiltonian with a renormalized velocity 
$v_F'=v_F\left(1+\frac{J_z}{\pi J}\right)$, and 
$H_J$ is a marginal current-current interaction.
Note that  $H_J$ can be incorporated into the free part,
\begin{equation}
{\cal L}=\frac{1}{\pi}\left(
\psi_L^\dagger\bar{\partial}\psi_L+\psi_R^\dagger\partial\psi_R\right)
+4\frac{g}{2\pi}J_LJ_R 
\rightarrow\frac{1}{2\pi r^2}\partial\varphi\bar{\partial}\varphi ,
\end{equation}
with dimensionless coupling 
$r^2=1/(1+4g) \sim 1-2J_z/\pi J$.\cite{Com2}

Let us first note that neither $H_J$ nor 
$H_b$ lifts the two-fold degeneracy at $\phi=1/2$,
hence causing no effects on the spectral flow. 
We thus concentrate on the behavior around $\phi=1$.
Recall that the effect due to $H_J$ can be incorporated in  
the free form
$E_{0J}'(Q_L,Q_R;N_L,N_R;\phi)=E_{0}(rQ_L,rQ_R;N_L,N_R; r \phi)$
in unit $2\pi v_F'/l$. 
So, $H_J$ only separates the primary states 
$|a \rangle$ and $|b \rangle$
from the descendant states  $|c \rangle$ and $|d \rangle$,
so that the state $|a \rangle$ can be 
raised up with smooth crossing with  $|b \rangle$ at 
$\phi =1$. This implies that the marginal operator $H_J$
does not affect the periodicity of the spectral flow,
i.e. the period is still $N$. 

We now look  at the effect of
$H_b$, which is an irrelevant operator for $J_z/J \le 1$
and vanishes at the massless fixed point after the
renormalization. Nevertheless, it can still affect the spectral flow
for the finite-size  system. 
 By recalling the bosonization rules for
fermion operators,
$\psi_{L}\rightarrow V_{Lr}\equiv {\rm e}^{{\rm i}r\varphi_L}$ and  
$\psi_{R}\rightarrow V_{R-r}\equiv {\rm e}^{-{\rm i}r\varphi_R}$,
the interaction $H_b$ is replaced by the sine-Gordon form,
\begin{equation}
H_b\rightarrow -\frac{g}{\pi}\int_0^l{\rm d}x
\cos 2r\varphi ,
\end{equation}
in a standard way, where $\varphi_L$ ($\varphi_R$) is
a left- (right-) moving component of $\varphi$, i.e., 
$\varphi =\varphi_L+\varphi_R$.
We now wish to evaluate the effect of $H_b$ on the 
spectrum at $\phi=1$ in the first order perturbation for the 
finite-size system with length $l$.
The two-fold degenerate primary states $|a \rangle$ and $|b \rangle$
can be  written as
\begin{equation}
|a\rangle =|V_{Lr}\rangle |V_{Rr}\rangle ,
\hskip 8mm
|b\rangle =|V_{L-r}\rangle |V_{R-r}\rangle ,
\end{equation}
where $|V_{L(R)r}\rangle$ denotes the primary state 
with dimension $r$. From the asymptotic behavior 
of the two- and three-point functions of 
the vertex operators,\cite{Car} we can see
$\langle r|V_{L(R) 2r} (x)|-r\rangle 
=\left(\frac{2\pi}{l}\right)^{2r^2}$.
The matrix element between these states
is thus calculated as
\begin{eqnarray}
\langle a|H_b|b\rangle &&=-\frac{g}{2\pi}\int_0^l{\rm d}x
\langle V_{Lr}|V_{L 2r} |V_{L-r}\rangle 
\langle V_{R r}|V_{R 2r} |V_{R -r}\rangle\nonumber\\
&&=-g\left(\frac{2\pi}{l}\right)^{4r^2-1} .
\end{eqnarray}
One can thus see from this expression 
that the twofold degenerate primary 
states are lifted by this irrelevant perturbation,
making the new states 
$|\pm\rangle\equiv (|a\rangle\pm |b\rangle )/\sqrt{2}$.
This level repulsion  produces a gap between them,
$\Delta
=2g\left(\frac{2\pi}{l}\right)^{4r^2-1}$.\cite{Com3}
Hence, with the increase of $\phi$ the 
initial ground state is continuously raised up 
to the excited states till $\phi=1$, then follows the lower branch
$|+\rangle$ smoothly,
 and finally returns to the absolute ground state at $\phi=2$.
We thus end up with 
the conclusion that the period of the spectral flow 
reduces from $N$  to 2 due to irrelevant interaction 
$H_b$, which reproduces the results of 
Sutherland and Shastry.\cite{SutSha}
An important point in the present analysis is that 
this periodicity is determined by 
the existence of irrelevant perturbations.
So, we can say that the period 2 for the spectral flow is 
rather common to  general (integrable and non-integrable)
spin chain  models 
as well as to interacting lattice fermion models in massless
phase, because these irrelevant interactions are 
naturally involved in ordinary lattice models.
Before closing this section, we wish to mention
the following point.  So far we have restricted ourselves 
to the antiferromagnetic cases.
In the case of the ferromagnetic interaction ($J_z<0$)
with massless excitations, 
it may not be easy to treat the spectral flow
directly by our approach, because 
bound states are formed during the process of the 
spectral flow, and their finite-size effects play a quite specific
role for the period of the ground state.\cite{YuFow,Sut,RES,FIO}
In particular, for special values of the coupling $J_z$
the period of the flow is completely modified, which
may not be treated by the present approach naively.
Field theoretic description of the ferromagnetic case
is still open and to be solved in the future study.

\section{Haldane-Shastry Spin Chain with $1/r^2$ Interaction}

It is now interesting to ask what happens for the 
spectral flow if we sweep away all the irrelevant interactions 
from the model. As already mentioned, the free fermion model is 
a typical example without such irrelevant perturbations, 
which has the period $N$ for $N$-site  systems.
\cite{SutSha}
We have another interesting 
model which is completely free from such irrelevant 
interactions, i.e., the
Haldane-Shastry (HS) spin model with $1/r^2$ interaction.
\cite{Hal2,Sha}
This Hamiltonian is known as the 
fixed-point Hamiltonian which describes ``free particles"
obeying fractional exclusion statistics  with statistical 
interaction parameter $g$.\cite{Hal} 
One would  naively expect that the period
of the spectral flow may be $N$, because there is no level-repulsion 
among the energy levels  due to irrelevant interactions. 
This question was actually raised in \cite{RES}, but has remained
still open. We wish to address this problem in the 
remainder of the paper. It will be shown that 
the period of the spectral flow 
in this model is naturally interpreted as $g$, 
{\it which directly reflects fractional exclusion statistics}.

Let us introduce the model Hamiltonian describing
the spin chain with $1/r^2$ exchange,\cite{Hal2,Sha}
\begin{eqnarray}
H&&=\frac{1}{2}\sum_{n=1}^N\sum_{n'=1}^{N-1}\sum_{l=-\infty}^\infty
\frac{1}{(n'+lN)^2}\nonumber\\
&&\times\left(
S_n^xS_{n+n'+lN}^x+S_n^yS_{n+n'+lN}^y+\Delta S_n^zS_{n+n'+lN}^z
\right) ,
\label{OriHam}
\end{eqnarray}
where we take the anisotropic parameter 
$\Delta =\frac{1}{2}g(g-1)$ with an even integer $g$.
It is known that $g$ is regarded as  
statistical interaction in ideal exclusion statistics.\cite{Hal}
Instead of periodic boundary conditions for the 
ordinary Haldane-Shastry model,\cite{Hal2,Sha} 
we impose twisted boundary condition (\ref{TwiBou})
in order to study the spectral flow. 
At $\phi=0$, the model is reduced to the ordinary 
Haldane-Shastry model.\cite{Hal2,Sha}

As has been shown in Ref.\cite{FukKaw},
the exact eigenstate of the twisted model can be obtained in the 
well-known Jastrow form, and its energy spectrum is correctly 
reproduced by the asymptotic Bethe ansatz.
The resulting expression is quite simple, which
is written down here. Namely, a certain series of 
the exact spectrum for the sector of 
$S_z=N/2-M$ is given by
$E_{\rm total}=(\frac{\pi}{N})^2(E(\phi )+e)$, where
$e=\frac{1}{6}(N^2-1)\left\{
\frac{1}{4}N\Delta +M(1-\Delta )\right\}$ and
\begin{equation}
E(\phi )=\sum_{i=1}^M\varepsilon (\widetilde k_i) .
\label{Ene}
\end{equation}
The single-particle dispersion relation in this expression is 
explicitly given by
\begin{equation}
\varepsilon (\widetilde k)=\left(2[\widetilde k]
-N+1\right)\widetilde k
-[\widetilde k]\left([\widetilde k]+1\right) ,
\label{Dis}
\end{equation}
where $[k]$ denotes  the Gauss symbol.
Here $\widetilde k$ is a pseudo-momentum including the 
effects of the twist, which is to be determined. 
Note that the effect of $1/r^2$ interaction is incorporated 
via the renormalization of $\widetilde k$ in eq.(\ref{Ene}), 
which is given by a solution to the Bethe equation,
\begin{equation}
\widetilde k_i=I_i+\phi +\frac{1}{2}(g-1)\sum_{j(\ne i)}{\rm sgn}
(\widetilde k_i-\widetilde k_j),
\label{ABA}
\end{equation}
where $I_i$ is an integer (or half integer) 
which specifies the energy spectrum.
We note that the key quantities to correctly 
follow the flow of the spectrum are the set of 
quantum numbers $\{ I_i \}$, and also the crystal momentum,
\begin{equation}
K(\phi )=\frac{2\pi M}{N}\sum_{i=1}^M\widetilde k_i .
\label{Mom}
\end{equation}

\subsection{Isotropic case}
By taking a suitable set of quantum numbers $\{ I_i \}$, 
we can determine the exact flow of the spectrum.
Let us first consider the isotropic case $\Delta=1$
($g=2$) by assuming that $N$ is a multiple of 4.\cite{FukKaw}
We see from (\ref{Ene}) that the spectral flow of the ground state
occurs continuously up to $\phi=1$
with the adiabatic increase of $\phi$,
by setting  $I_i =\left(M-1\right)/2+i$ with $i=1,2,\cdots,M$.
Although we encounter a cusp structure at $\phi=1$
in (\ref{Ene}), we may pass through it by keeping the
quantum numbers $\{I_i\}$ unchanged modulo the 
periodicity of them in (\ref{ABA}). Based on these observations, 
we obtain the natural spectral flow of the 
ground state for the system with the statistical interaction $g=2$,
resulting in the expression,
\begin{equation}
E(\phi )=\left\{
\begin{array}{ll}\frac{N}{2}\phi
-\frac{1}{12}N(N^2+2),& \quad {\rm for}\quad 0\le\phi\le 1\\
-\frac{N}{2}\phi
-\frac{1}{12}N(N^2-10),& \quad {\rm for}\quad 1\le\phi\le 2
\end{array}\right.
\label{GroSpe2}
\end{equation}
Therefore the period of the spectral flow is 
2, which may have a different origin from  that for the 
nearest-neighbor Heisenberg model.
We will indeed explain below that the obtained spectral flow 
actually realizes the {\it motif picture} of fractional exclusion 
statistics.  Since the above exact solution 
produces only a certain series of the exact spectrum, 
we have complimentarily calculated the exact spectral flow  for the 
finite system numerically. In Fig.\ref{Flow80}, the results for the 
$N=8$ system with  $\Delta=1$ ($g=2$) are shown.
It is seen that there are not 
level repulsions at any level crossing points,
which we have discussed for the nearest-neighbor model in
the previous section. They are replaced by 
characteristic cusp structures, which are indeed 
observed in the exact result of (\ref{GroSpe2}).
One can easily trace the spectral flow of the ground state 
in Fig.\ref{Flow80} according to (\ref{GroSpe2}).

\subsection{Anisotropic case}
In contrast to the isotropic case, it may not be straightforward
to discuss the spectral flow in the anisotropic case.
It is indeed not trivial to figure out what kind of the 
ground state is realized in thermodynamic limit.
Nevertheless,
we find that the present approach can be still applied
to discuss properties of 
the finite system. For example, the spectrum shown 
in Fig.\ref{Flow82}, which is numerically calculated
for the $N=8$ system with $\Delta=6$ ($g=4$), can be 
well described by (\ref{ABA}). Namely, 
similar analyses to the isotropic case enable us to 
write down the analytic expression for the natural 
spectral flow as
\begin{equation}
E(\phi )=\left\{
\begin{array}{ll}
\frac{N}{4}\phi
-\frac{1}{24}N(N^2+8),& \quad {\rm for}\quad 0\le\phi\le 1\\
\frac{3N}{4}\phi
-\frac{1}{24}N(N^2+20),& \quad {\rm for}\quad 1\le\phi\le 2\\
-\frac{3N}{4}\phi
-\frac{1}{24}N(N^2-58),& \quad {\rm for}\quad 2\le\phi\le 3\\
-\frac{N}{4}\phi
-\frac{1}{24}N(N^2-16),& \quad {\rm for}\quad 3\le\phi\le 4
\end{array}\right.
\label{GroSpe4}
\end{equation}
For example, it is seen from the momentum 
conservation (\ref{Mom}) that starting from the ground state 
at $\phi=0$, we should 
follow the upper branch of the flow (ii)  after 
the level crossing point at $\phi=1$ in Fig.\ref{Flow82}.
The expression (\ref{GroSpe4})
is obtained  for the sector with the total spin $S_z=2$,
which is confirmed to be indeed the ground state 
for the $N=8$ system. Thus, the period of the spectral flow 
for the ground state is regarded as 4, 
reflecting the statistical interaction $g=4$.
 
\subsection{Motif picture}
We now discuss the origin of the period 
for the Haldane-Shastry model. Let us recall again that  
the Haldane-Shastry model describes the fixed-point 
Hamiltonian without irrelevant perturbations,
which is indeed consistent with the above results 
that there is no level repulsions at any level-crossing points.
So, in this ideal situation, 
it may be expected that one can see characteristic properties of
fractional (exclusion) statistics in the spectral flow.\cite{Hal}
To explain this in an intuitive way, it is convenient to  
use the notion of {\it motif},\cite{HHTBP} 
which is a fermionic occupation-number classification in the
momentum space. 
We describe the behavior of the spectral flow 
in this language.  The statistical interaction $g>1$ means that the 
Pauli-principle acts stronger than the fermionic case due
to the $1/r^2$ interaction, and as a consequence 
the spacing of the occupied states should be 
enlarged  $g$-times larger than that of free fermions.
This naturally leads us to the description by motif. 
To see the motif description clearly, we first
describe the free system in this language.
In this case, the Pauli principle controls the 
occupation of levels in the momentum space.
Let us denote an occupied (unoccupied) state 
by 1 (0). Since there is no interaction among particles, 
the ground state motif 
is given by $000\cdots 0011\cdots 1100\cdots 000$ which consists 
of $N/2$ 0's and $N/2$ 1's.
Note that adding a unit flux $\phi$ shifts all 1's 
uniformly to the right by one step. 
So, it is easily seen  that this motif returns to 
the original one after adding the $N$ flux. 
Therefore, we can see that the period for the 
free fermion system is $N$, being consistent with the 
exact results mentioned before.

Let us now consider the present $1/r^2$ model with
$g=2$.  In this case, the corresponding motif is 
constructed by all sequences of 0 and 1 not containing one or
more consecutive 1's, which reflect the repulsive interaction 
with $g=2$. In the same way above, since
adding a unit $\phi$ shifts 1's to the right by one step, 
the ground state motif behaves as
\begin{equation}
01010\cdots 01010\stackrel{\delta\phi =1}{\longrightarrow}
00101\cdots 10101
\stackrel{\delta\phi =1}{\longrightarrow}01010\cdots 01010 .
\end{equation}
Observing this, we see that the period of the spectral flow is
indeed 2, which exactly reproduces (\ref{GroSpe2}).
On the other hand, the motif for the ground state 
of the finite system shown in Fig.\ref{Flow82} 
($g=4$, $N=8$) is $001000100\cdots 001000100$.
Similarly to the $g=2$ case, one easily finds that the
motif has a period 4 , reflecting the statistical interaction $g=4$.
So, we can naturally interpret the period observed for 
the Haldane-Shastry model in terms of the fractional
exclusion statistics.


\section{Summary}

We have studied the adiabatic flow of the energy 
spectrum for spin chains.
By using effective field theory for 
the $XXZ$ Heisenberg model, we have clarified that 
the period $2$ obtained by Sutherland and Shastry 
arises from leading irrelevant perturbations in the 
model. This implies that the period $2$ is quite common 
to many 1D interacting quantum systems at half-filling
(or spin systems with zero magnetic field), because 
ordinary quantum models naturally involve such perturbations.
On the other hand, the quantum models with $1/r^2$
interaction has a nature of the fixed point Hamiltonian which is
free from such irrelevant interactions.
We have demonstrated that in this ideal system
we can naturally see the statistical interaction in the period 
of the spectral flow.

\acknowledgements{}
We would like to thank Y. Kuramoto, Y. Kato,
S. Fujimoto, T. Yamamoto and H. Awata for valuable discussions.
This work is supported by the Grant-in-Aid from the Ministry of
Education, Science and Culture, Japan.

\appendix
\section{Parity Effect on Jordan-Wigner Fermions}
In this appendix, we briefly 
discuss the parity effect on the ground state properties
for Jordan-Wigner fermions.  First let us recall 
the well-known fact that the ground state 
for spinless fermion systems at $\phi=0$ is 
unique (doubly degenerate) according to whether the 
number of fermions is odd (even).  So, if   
we introduce the external gauge field or 
alternatively twist boundary conditions, 
the system shows diamagnetic (paramagnetic) 
response depending on the parity with respect to the number of 
particles, i.e. the ground state energy 
first increases (decreases) with the increase of $\phi$.
In contrast to this, we show below that 
the ground state at $\phi=0$
for the present system is always unique 
and the response to the gauge field is diamagnetic.

First, we rewrite the boundary condition (\ref{TwiBou})
in terms of Jordan-Wigner fermions,
\begin{equation}
\psi_{N+1}=\psi_1{\rm e}^{2\pi{\rm i}(\phi +1/2-{\cal N}/2)}
\equiv\psi_1{\rm e}^{2\pi{\rm i}(\phi +\phi_0)},
\label{FerTwi}
\end{equation}
where ${\cal N}\equiv\sum_{j=1}^{N}n_j$ and 
$\phi_0\equiv (M-1)/2$ with $M$ being the number of fermions.
Note that the extra phase $\phi_0$, originating from 
the minus sign in $\psi_{N}^\dagger {\rm e}^{-{\rm i}\pi{\cal N}}\psi_1
=-\psi_{N}^\dagger\psi_1{\rm e}^{-{\rm i}\pi{\cal N}}$,
reflects the fact that the spin system is equivalent 
to the hard-core bosons. 
For even $N$, we can see that $\phi_0=1/2~(0)$ for $N/2=$ even (odd)
for the half filling $M=N/2$. This phase factor $\phi_0$
distinguishes the present system from the 
ordinary fermion system, and  plays a role 
to determine the ground state uniquely.

To confirm the above statement explicitly,
we express the free  Hamiltonian (\ref{HamFre}) 
in the diagonal form in  Fourier space.  The corresponding  
one-particle dispersion is given by
$\epsilon (k)\equiv -J\cos
2\pi (k+\phi +\phi_0)/N\equiv -J\cos (ap)$.  
Here the quantized momentum $p$ is defined by 
$p=2\pi (k+\phi +\phi_0)/l\equiv 2\pi n/l$.
Note that the twist angle shifts the momentum and plays 
a role of gauge potential.
At $\phi =0$, $n\in {\bf Z}+\phi_0$, namely, 
$n\in {\bf Z}+1/2~({\bf Z})$
for $N/2\equiv M$ even (odd). 
The ground state configuration is then 
$n=-(M-1)/2,-(M-1)/2+1,\cdots, (M-1)/2$
both for cases $N/2=$even and odd. 
Therefore, we can see that the ground state is always unique 
irrespective of the parity of the particle number.
This is due to the extra phase $\phi_0$ in eq.(\ref{FerTwi}).

Low energy excitations near the ground state are classified by 
the momentum $ap=a(\pm p_F+p')$, where 
$ap'=2\pi (k'+\phi +1/2)/N$ describes the 
excitation near the Fermi points. Note that 
the extra factor $1/2$ is always accompanied 
in this expression, so that it is convenient to 
introduce $ \tphi\equiv\phi +1/2$ to simplify
the expressions in the text. 

\begin{figure}[h]
\caption{Schematic diagram for spectral flow in
 the non-interacting case. 
It is described how the ground state configuration at
$\phi=0$ evolves to the corresponding configuration
 at $\phi=1$ (occupied states for left- and right-going
sectors are denoted by dots). 
It can be seen that 
$|0,0;0,0;\phi=1\rangle=|-1,1;0,0;\phi=0\rangle$.
Therefore, if we measure the fermion numbers $Q_L$ and $Q_R$
with respect to 
the vacuum state defined in the text, 
there occurs a pair creation at $\phi=1/2$.
Namely, $Q_R-Q_L$ changes by 2 during this process with 
keeping $Q_L+Q_R=0$.}
\label{OnePF}
\end{figure}

\begin{figure}[h]
\caption{Exact spectral flow for the $S^z=0$ sector of 
the $N=8$ system with  $\Delta =1 ~(g=2)$.
Lower 10 levels are described.}
\label{Flow80}
\end{figure}

\begin{figure}[h]
\caption{Exact spectral flow for the $S^z=2$ sector of 
the $N=8$ system with $\Delta =6~(g=4)$.
 Lower 10 levels are described.
The spectral flow of the ground state occurs like 
i$\rightarrow$ii$\rightarrow$iii$\rightarrow$iv.
Note that we have folded the picture for $ 0 \le \phi \le 4$
with respect to $\phi=2$ for simplicity.}
\label{Flow82}
\end{figure}

\begin{references}
\bibitem[*]{JSPS} JSPS Research Fellow. 
email address: fukui@yukawa.kyoto-u.ac.jp
\bibitem{SutSha} B. Sutherland and B.S. Shastry: Phys. Rev. Lett. 
{\bf 65} (1990) 1833.
\bibitem{YuFow} N. Yu and M. Fowler: Phys. Rev. {\bf 46} (1992) 14583.
\bibitem{Sut} B. Sutherland: Phys. Rev. Lett. {\bf 74} (1995) 816.
\bibitem{RES} R.A. R\"omer, H.-P. Eckle and B. Sutherland:
Phys. Rev. {\bf B52} (1995) 1656.
\bibitem{FIO} N. Fumita, H. Itoyama and T. Oota: preprint.
\bibitem{KorWu} V.E. Korepin and A.C.T. Wu: Int. J. Mod. Phys.
{\bf B5} (1991) 497.
\bibitem{Aoki} K. Kusakabe and H. Aoki: ISSP preprint.
\bibitem{Hal} F. D. M. Haldane: Phys. Rev. Lett. {\bf 67} (1991) 937.
\bibitem{Aff} I. Affleck: in Les Houches 1988, eds. E. Br\'ezin and
J. Zinn-Justin.
\bibitem{Com1}To see the spectral flow clearly, we have used
the quantum numbers $Q_L$ and $Q_R$ which classify the primary 
states at $\phi=0$.
\bibitem{Car} J.L. Cardy: Nucl Phys. {\bf 270} (1986) 186.
\bibitem{Com2}
This coupling is correct for small $J_z$.
For an arbitrary interaction for massless cases,
Bethe ansatz solution gives the exact 
value $r^2=\pi /2(\pi -\mu )$
where $\mu$ is defined by $J_z/J=\cos\mu$.
\bibitem{Com3}Note that these states $|\pm\rangle$
have the momentum $\pi$ and correspond to the lowest
states of the des Cloizeaux-Gaudin dispersion at the momentum $\pi$.
See J. des Cloizeaux and Gaudin: J. Math. Phys. {\bf 7}	(1966) 1384.
\bibitem{Hal2} F.D.M. Haldane: Phys. Rev. Lett. {\bf 60} (1988) 635.
\bibitem{Sha} B.S. Shastry: Phys, Rev. Lett. {\bf 60} (1988) 639.
\bibitem{FukKaw} T. Fukui and N. Kawakami: Phys. Rev. Lett. 
{\bf 76} (1996) 4242. 
In this paper, the spectral flow for the  case 
of $g=2$ is briefly discussed.
\bibitem{FUKUI} T. Fukui and N. Kawakami: J. Phys. {\bf A28} (1995) 6027.
\bibitem{HHTBP} F.D.M. Haldane, Z.N.C. Ha, J.C. Talstra, D. Bernard
and V. Pasquier: Phys. Rev. Lett. {\bf 69} (1992) 2021.
\end{references}
\end{document}